\documentclass[aps,prl,twocolumn,groupedaddress,showpacs]{revtex4}
\usepackage{graphicx}% Include figure files
\usepackage[dvips]{epsfig}% Include figure files
\usepackage{dcolumn}% Align table columns on decimal point
\usepackage{bm}% bold math
\usepackage{amssymb}
\usepackage{amsmath}

\begin{document}

\title{Switching rates of multi-step reactions}

\author{Carlos Escudero$^\dag$ and Alex Kamenev$^\ddag$}

\affiliation{$\dag$ Instituto de Matem\'{a}ticas y F\'{\i}sica Fundamental, Consejo Superior de Investigaciones Cient\'{\i}ficas, C/ Serrano 123, 28006 Madrid, Spain \\
$\ddag$ School of Physics\&Astronomy, University of Minnesota, 116 Church St. Minneapolis, MN 55455, USA}

\begin{abstract}
We consider a switching rate of a meta-stable reaction scheme, which  includes reactions with  arbitrary steps, e.g. $kA\to(k+r)A$. Employing   WKB approximation, controlled by a large system size, we evaluate  both the exponent and the pre-exponential factor for the rate. The results are illustrated  on a number of examples.
\end{abstract}

\pacs{05.40.-a,02.50.Ga,64.60.My,82.20.-w}

\maketitle

Since the celebrated work of Kramers on reaction-rate theory \cite{kramers}, much effort has been devoted to extending and generalizing his results, see Ref.~[\onlinecite{hanggi}] for a review.
%This is due, doubtless, to the practical omnipresence of the problem of escape from metastable states in science.
Applications of this theory can be found in fields as diverse as
high energy physics, nucleation, chemical kinetics, electric
transport, diffusion in solids and population dynamics among many
others. In this work we consider  a switching rate in a generic
reaction scheme, which admits more than one (quasi)stationary state.

A particular case of  single-step reactions allows for an exact solution and is well-studied in the literature \cite{gardiner,doering}. We thus  concentrate on  generic multi-step reactions. Although an exact solution is not known, a substantial progress may be achieved by adopting an analog of the  quantum mechanical WKB approximation \cite{kubo,dykman1,elgart}, controlled by a large system size.  With an exponential accuracy it gives the switching rate as an exponentiated action of an auxiliary mechanical problem. Evaluation of the pre-exponential factor requires a matching of the quasi-stationary distribution (QSD) function, found in the WKB framework, with the constant current ''behind the barrier'' solution \cite{kramers,dykman}. The first consistent application of this strategy to a model reaction scheme was presented recently  by  Meerson and Sasorov \cite{meerson}. Here we generalize their approach to an
arbitrary scheme with metastable states.

%{\em Mathematical formulation.}
Consider a generic multi-step reaction scheme, where a state with
$n$ particles may be transformed into a state with $n+r$ particles  with the
rate $W_r(n)$. Here $r$ is a set of integers not necessary equal $\pm
1$.  The corresponding Master equation for the probability distribution $P_n(t)$ is
\begin{eqnarray}\nonumber
    \partial_t P_n(t) &=&\sum_r\left[ W_r(n-r)P_{n-r}(t) -W_r(n)P_n(t)\right]\\ &=&
\sum_r\left(e^{-r\partial_n}-1\right) W_r(n)P_n(t)\,.
\label{master}
\end{eqnarray}
We  focus on reactions which admit a QSD centered at $n=n_0$ and
an unstable equilibrium  (saddle point) at $n=n_s$. For definiteness we  assume that $n_0<n_s$. We  also assume that both
$n_0$ and $n_s$ scale in the same way with a large parameter $N\gg 1$, hereafter referred to as the system size, i.e. $n_{0,s}\sim N$. It is then convenient to pass to a scaling variable $q=n/N$ and separate the leading and the first subleading orders in $N$ in the corresponding reaction rates
\begin{equation}\label{W}
    W_r(n)=Nw_r(q) + u_r(q) + O(1/N)\,; \quad q=n/N.
\end{equation}

We  seek for QSD  in the form $P_n(t)= \pi(n) e^{-E_0t}$, where $E_0=1/\tau$ is an exponentially small eigenvalue of the Master equation. In the rescaled coordinate the corresponding eigenvector may be sought in the WKB form
\begin{equation}\label{pi}
    \pi(q)= \exp\{-NS(q)-S_1(q)\}\, .
\end{equation}
Substituting this form in the Master equation (\ref{master}) and keeping terms up to the order of  $O(1)$, one finds
\begin{eqnarray}
\nonumber
0 &=& \sum_r \left(Nw_r +  u_r\right)\\ &\times &
\left( e^{ rS'}\left[1-\frac{r^2}{2N}\,S'' + \frac{r}{N}S_1'\,
-\frac{r}{N}\, \frac{w_r'}{w_r}\right]-1 \right) ,
\label{master1}
\end{eqnarray}
where the primes denote derivatives with respect to rescaled reaction coordinate $q$. We have also took into account that the
eigenvalue $E_0$ is expected to be exponentially small in $N$ (see below) and thus may be omitted.

In the order $N$ this equation acquires a form of the stationary
Hamilton-Jacobi equation $H(q,S')=0$, where the effective
classical Hamiltonian takes the form \cite{dykman1,dykman}
\begin{equation}\label{H}
    H(q,p)= \sum_r w_r(q) \left(e^{ rp} -1 \right)\, ,
\end{equation}
and we have denoted $S'=p$. Therefore to the order $N$ the problem is reduced to finding zero energy trajectories $p=p(q)$, such that
$H(q,p(q))=0$, of a corresponding  ''mechanical'' problem.
%In mechanical terms we would say that the ''action'' $S$ is obtained as the primitive of the ''momentum'' $p$: $S(q)=\int^q p(x)dx$.

\begin{figure}
\begin{center}
\psfig{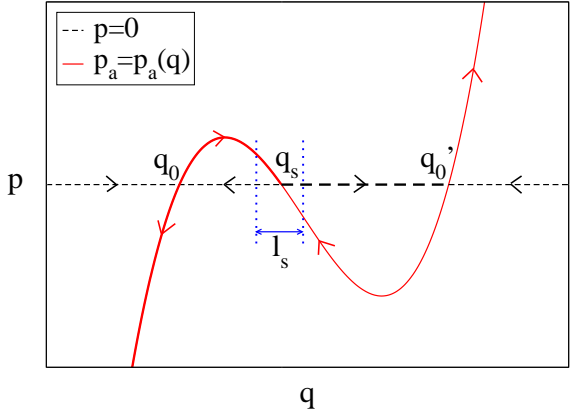}
\end{center}
\caption{(Color online) Phase portrait of a typical bistable reaction. The  dashed line is the relaxation trajectory $p=0$, the solid line is the activation trajectory $p_a=p_a(q)$. These zero-energy curves intersect at the metastable  points $q_0$, $q_0'$ and the saddle point $q_s$. The arrows show direction of motion according to the  Hamilton equations. The matching of activation and relaxation solutions takes place in a narrow region of the width $l_s\sim N^{-1/2}$ around the saddle point.} \label{fig1}
\end{figure}

The phase portrait of a typical bistable reaction is plotted in Fig.
\ref{fig1}. There are at least two appropriate zero energy
trajectories: the {\em relaxation} trajectory $p=0$ and the {\em
activation} trajectory  $p=p_a(q)$, see Fig. \ref{fig1}. The
classical equation of motion along the relaxation  path $\dot q =
H_p(q,0) =\sum_r r w_r(q)$ is nothing but the mean-field rate
equation for our reaction scheme. According to our assumptions it
admits  stationary states $q_{0,s}=n_{0,s}/N$, where
$H_p(q_{0,s},0)=0$ (other stationary states are possible, e.g.
$q_{0}'$ see Fig.~\ref{fig1}). Those are the points, where the
activation trajectory $p_a(q)$ crosses the relaxation one $p=0$ and
thus $p_a(q_{0,s})=0$.

To escape from a metastable state centered around $q_0$ the system must evolve along the activation trajectory, Fig.~\ref{fig1}. The QSD
is given by Eq.~(\ref{pi}), where $S(q)$ and $S_1(q)$ are determined by the order $N$ and order $1$ terms in  Eq.~(\ref{master1}) correspondingly. They lead to
\begin{eqnarray}
  S(q)\!\! &=&\!\! \int^q\!\! dq\, p_a(q)\, ; \label{S} \\
  S_1(q)\!\! &=& \!\! \int^q\! \! dq \, \frac{p_a'H_{pp}\! +\! 2 H_{pq}\! -\! 2\sum_r u_r (e^{rp_a}-1) }{2H_p},
  \label{S1}
\end{eqnarray}
where derivatives of the Hamiltonian are evaluated along the activation path, e.g. $H_{pq}=\sum_r re^{rp_a(q)}w'_r(q)$, {\em etc} and
$p'_a=S''$. Equations (\ref{pi}), (\ref{S}), (\ref{S1}) determine  QSD up to a multiplicative constant. To find the latter, one needs to
match the QSD with the constant current solution on the other side of the saddle point $q_s$ \cite{kramers,dykman,meerson}.

At $q>q_s$ the system evolves along the relaxation trajectory $p=0$, Fig.~\ref{fig1}, and therefore $S\equiv 0$. Solving Eq.~(\ref{master1}) for $S_1$, one finds
\begin{equation}\label{as}
    \pi(q) = J/H_p(q,0)\,,
\end{equation}
where $J$ is an integration constant given by the current out of QSD. Indeed,  the Master equation (\ref{master}), having the structure of the continuity relation, in a vicinity of the relaxation trajectory $p=0$ acquires a form
\begin{eqnarray}
\label{cont}
\partial_t P(q,t) =  -\partial_q \big[H_p(q,0)P(q,t) + O(1/N)\big] \,.
\label{continuity}
\end{eqnarray}
Therefore the relaxation limit (\ref{as}) of  QSD $P(q,t)=\pi(q) e^{-E_0t}$ is nothing but a constant current, $J$, solution of the Master equation (where we have again neglected the exponentially small eigenvalue $E_0$ on the l.h.s.). On the other hand, integrating the continuity relation (\ref{cont})  throughout the region of support of QSD and assuming that escape takes place {\em only} through the saddle point $q_s$ \cite{foot3}, one finds
\begin{equation}
\label{norm}
E_0 \int\!\! \pi(q) dq = J\, .
\end{equation}

Finally to establish  relation between the activation solution, Eqs.~(\ref{pi}), (\ref{S}), (\ref{S1}), at $q<q_s$ and the relaxation one,  Eq.~(\ref{as}),  at
$q>q_s$, one needs to consider Master equation in an immediate vicinity of the saddle $q_s$ \cite{meerson}.
Expanding the r.h.s. of Eq.~(\ref{master}) to the second derivative, one finds for the current:
\begin{equation}\label{Kampen1}
    \big[H_{pq}(q_s,0)\big] (q-q_s) \pi(q)  - \frac{H_{pp}(q_s,0)}{2N}\,   \partial_q \pi(q) = J\,,
\end{equation}
where we have used the fact that at the saddle point $H_p(q_s,0)=\sum_r r w_r(q_s)=0$.
Solution of Eq.~(\ref{Kampen1}) with a proper asymptotic behavior has the following  form
$\pi(q)=(2NJ/H_{pp})\, e^{(q-q_s)^2/l_s^2} \int_{q-q_s}^\infty dq\, e^{ -(q-q_s)^2/l_s^2}$, where
$l_s^2=H_{pp}(q_s,0)/NH_{pq}(q_s,0)$.
Indeed, away from the saddle point $q_s$ it possesses  the following asymptotics:
\begin{eqnarray}
    \pi(q)=
    \left\{ \begin{array}{ll}
    \frac{J}{(q-q_s) H_{pq}};  &\quad\quad q-q_s\gg l_s\,, \\  \\
    \frac{2NJl_s\sqrt{\pi}}{H_{pp}}\, \, e^{(q-q_s)^2/l_s^2}; \,&\quad \quad q_s-q\gg l_s\,.
    \end{array} \right.
\label{asymptotic}
\end{eqnarray}
The first line matches with the relaxation solution (\ref{as}) at $q\approx q_s$, as it should. The second line is to be matched with the activation
solution Eqs.~(\ref{pi}), (\ref{S}), (\ref{S1}), which in the vicinity of $q=q_s$ takes the form
\begin{equation}\label{pif}
    \pi(q)=e^{-NS(q_s)-S_1(q_s)}\, e^{-N(q-q_s)^2S''(q_s)/2}.
\end{equation}
To relate the $q$-dependent exponential factors here and in the second line of Eq.~(\ref{asymptotic}) one may differentiate the identity
$H(q,p_a(q))=0$ over $q$ to find
\begin{equation}
\label{identity}
H_q + H_p p_a'=0\,;\quad \quad
H_{qq}+  H_p p_a'' + (H_{pp}p_a' + 2H_{pq})p_a'= 0\,.
\end{equation}
Employing that $p_a'=S''$ and $H(q,0)=H_p(q_{0,s},0)=0$,  one finds
\begin{equation}
\label{S-double-prime}
S''(q_{0,s}) = - \frac{2H_{pq}(q_{0,s},0)}{H_{pp}(q_{0,s},0)}=- \frac{2\sum_r rw_r'(q_{0,s})}{\sum_r r^2 w_r(q_{0,s})}
\end{equation}
and therefore $S''(q_s)=-2/Nl_s^2$.
This establishes equality of the exponential factors in Eqs.~(\ref{asymptotic}) and (\ref{pif}).
Comparing the pre-exponential coefficients one finds for the escape current:
\begin{equation}\label{J}
    J=\frac{H_{pp}(q_s,0)}{2} \sqrt{\frac{|S''(q_s)|}{2\pi N}} \,\, e^{-NS(q_s)-S_1(q_s)}\, .
\end{equation}

One can employ now the normalization   condition (\ref{norm}) to
find the escape rate $E_0=1/\tau$. To this end we notice that the
bulk of the QSD is centered around $q_0$ and approximate the
integral by the Gaussian one. As a result one finds for the escape
time
\begin{equation}\label{tau}
    \tau = \frac{4\pi}{H_{pp}(q_s,0)}\,\frac{e^{S_1(q_s)-S_1(q_0)}}{\sqrt{|S''(q_s)|S''(q_0)}} \,\, e^{N[S(q_s)-S(q_0)]}\, ,
\end{equation}
where $S(q_s)-S(q_0)$ and $S_1(q_s)-S_1(q_0)$ are fully determined by Eqs.~(\ref{S}) and (\ref{S1}). It is important
to mention that the corresponding integrals are free of singularities and can be straightforwardly evaluated for any
given reaction scheme. Equation (\ref{tau}) is a main result of the present letter.

For analytically tractable examples it is useful to notice that, with the help of identities (\ref{identity}) one may partially
integrate Eq.~(\ref{S1}) to obtain
\begin{eqnarray}\label{S1new}
S_1(q_s)-S_1(q_0)&=&\ln \sqrt{\frac{S''(q_0)}{|S''(q_s)|}}\,\, +\, \Delta \,;\\
\Delta &=& \int\limits_{q_0}^{q_s}\!\! dq\,
\left[\frac{H_{qq}}{2H_q} -\frac{\sum_r u_r (e^{rp_a}-1) }{H_p} \right].
\nonumber
\end{eqnarray}
Employing Eq.~(\ref{S-double-prime}), one may somewhat simplify  Eq.~(\ref{tau}) to cast it in the following form
 \begin{equation}\label{tau1}
    \tau = \frac{2\pi\, e^{\Delta}}{H_{pq}(q_s,0)}\, \,\, e^{N[S(q_s) -S(q_0)]}\,.
 \end{equation}
Below we illustrate  usefulness of Eqs.~(\ref{tau}) and (\ref{tau1}) on a few  examples.

{\em $r_1$--$r_2$ reactions.} Consider a reaction scheme, where the step variable $r$ may acquire only two values $r_1$ and $r_2$.   The corresponding reaction  rates are $W_{r_{1,2}}(n)=Nw_{r_{1,2}}(q)$, where we have omitted possible subleading terms $u_{r_{1,2}}$ for brevity. The Hamiltonian takes the form
\begin{equation}\label{two-step}
    H(q,p)=w_{r_1}(q)(e^{r_1 p}-1)+w_{r_2}(q)(e^{r_2 p}-1)\,,
\end{equation}
and the activation trajectory is given by the solution of the following algebraic equation for $e^{p_a}$
\begin{equation}
\label{two-step1}
\frac{e^{r_1 p_a(q)}-1}{e^{r_2 p_a(q)}-1} = - \frac{w_{r_2}(q)}{w_{r_1}(q)} \,.
\end{equation}
As a result, the following identity holds along  the activation trajectory:
$$
\frac{H_{qq}}{H_q}=\frac{w_{r_1}''(q)(e^{r_1 p_a}-1)+w_{r_2}''(q)(e^{r_2 p_a}-1)}{w_{r_1}'(q)(e^{r_1 p_a}-1)+w_{r_2}'(q)(e^{r_2 p_a}-1)} $$
$$
=\frac{w_{r_1}w_{r_2}''-w_{r_1}''w_{r_2}}{w_{r_1}w_{r_2}'-w_{r_1}'w_{r_2}}=\frac{d}{dq}\mathrm{ln}(w_{r_1}w_{r_2}'-w_{r_1}'
w_{r_2})\, .
$$
The fixed points satisfy: $r_1w_{r_1}(q_{0,s})= -
r_2w_{r_2}(q_{0,s})$, while $H_{pq}(q_{0,s},0)= r_1
w_{r_1}'(q_{0,s})+r_2 w_{r_2}'(q_{0,s})$. Employing
Eqs.~(\ref{S1new}) and (\ref{tau1}), one finds for the switching
time
 \begin{eqnarray}
% \nonumber
\tau = 2\pi \sqrt{\left|\frac{w_{r_1}(q_s)}{w_{r_1}(q_0)}\right|}\,\, \frac{e^{N[S(q_s)-S(q_0)]}}{\sqrt{|H_{pq}(q_s,0)H_{pq}(q_0,0)|}}\, ,
%\\ \times \exp\left[-\int^{q_s}_{q_0} \frac{\sum_r u_r(x) (e^{rp_a(x)}-1) }{H_p(x,p_a(x))} \,\, dx \, \right] \,,
\label{tauts}
 \end{eqnarray}
where $w_{r_1}(q_s)/w_{r_1}(q_0)=w_{r_2}(q_s)/w_{r_2}(q_0)$  and the action is given by Eq.~(\ref{S}).

In a particular case of {\em single-step} reactions, $r_{1,2}=\pm 1$, Eq.~(\ref{two-step1}) may be solved explicitly, $e^{p_a(q)}= w_-(q)/w_+(q)$. The fixed points are given by $w_+(q_{0,s})= w_-(q_{0,s})$ and  according to Eq.~(\ref{S-double-prime}) $H_{pq}(q_{0,s},0) =- S''(q_{0,s}) w_+(q_{0,s})$.
Employing Eq.~(\ref{tauts}), the switching rate of the single-step reaction schemes may be written as
\begin{equation}\label{tauss2}
    \tau = \frac{2\pi}{w_\pm(q_0)}\, \frac{e^{-\int_{q_0}^{q_s}dq\left(\frac{u_+}{w_+} -
    \frac{u_-}{w_-}\right)  }} { \sqrt{|S''(q_s)|S''(q_0)} }  \,\, e^{N[S(q_s) -S(q_0)]}\,,
 \end{equation}
where
\begin{equation}\label{single-step-action}
    S(q_s) -S(q_0)=\int_{q_0}^{q_s}\!\! dq \, \ln\big( w_-(q)/w_+(q)\big)\,
\end{equation}
and we have included subleading terms in the rates $u_\pm (q)$, according to Eq.~(\ref{S1new}), [\onlinecite{foot1}]. In a particular case of reaction rates having only leading terms ($u_\pm =0$) Eq.~(\ref{tauss2}) coincides with results of Doering {\em et al.} [\onlinecite{doering}], who have shown it to be the large  $N$ asymptotic of the exact result for the single-step reactions \cite{gardiner}. In general, the $u_r$ terms can substantially  modify the prefactor \cite{meerson} (see below).

 {\em Demographic explosion.} Consider a single-step model \cite{elgart,meerson}  $A\rightleftarrows \emptyset$ with the relative rates $1$ and $N(1-\delta^2)/2$, where $0<\delta<1$, and $2A\to 3A$ with the relative rate $1/N$.  The corresponding transition rates are
$$
W_-(n) = n\,; \quad W_+(n) = \frac{N(1-\delta^2)}{2} + \frac{n(n-1)}{2N}\, .
$$
The rescaled rates are $w_-=q\,$;  $w_+=(1-\delta^2+q^2)/2$ while $u_-=0$ and $u_+=-q/2$
and the two rescaled fixed points are $q_{0,s}=1\mp\delta$.
Employing Eq.~(\ref{tauss2}), one finds for the escape time from the metastable state centered at $n=N(1-\delta)$ towards $n\to \infty$
 \begin{equation}
    \tau =
 %\frac{2\pi}{q_s-1 }  \sqrt{\left|\frac{q_s^2-\gamma} {q_0^2-\gamma} \right|}     \, e^{\int_{q_0}^{q_s}\frac{q dq}{q^2+\gamma}}\,\, e^{N\Delta   S}=
    \frac{2\pi}{\delta }  \frac{1+\delta} {1-\delta}  \,\, e^{N[S(1+\delta) -S(1-\delta)]} \,,
\label{tauss3}
 \end{equation}
in a perfect agreement with Meerson and Sasorov recent result [\onlinecite{meerson}]. This example is specially interesting because it shows the importance of the subleading terms $u_r$. Disregarding these terms, one obtains a prefactor proportional to $(1-\delta)^{-1/2}$ instead of the correct one $(1-\delta)^{-1}$. This constitutes an arbitrarily large error in the limit $\delta \to 1$, where the action $S(2)-S(0)$ remains bounded.

{\em Fokker-Planck Hamiltonian.} Consider a dissipative particle
under an  influence of a multiplicative Gaussian noise (understood in the sense of It\^{o} \cite{gardiner}). The corresponding
Fokker-Planck equation is $\dot P=\hat H P$, where
\begin{equation}\label{FP}
    \hat H(q,\hat p) = \hat p^2 D(q) - \hat p\, V'(q)\,,
\end{equation}
here $D(q) > 0$ is a coordinate-dependent diffusion coefficient
and $\hat p=-\partial_q$. Since this is a normally ordered
operator, cf. Eq.~(\ref{master}), one may employ the theory
developed above. Following WKB approximation one  substitutes
$\hat p\to p$ and employs Eq.~(\ref{tau1}). The stationary points
are defined by the condition $V'(q_{0,s})=0$ and the activation
trajectory  is given by $p_a(q)=V'(q)/D(q)$. As a result
$S(q_s)-S(q_0)=\int_{q_0}^{q_s}\!dq\, V'(q)/D(q)$ and
$H_{pq}(q_s,0)=-V''(q_s)>0$. There are no subleading terms here,
$u_r=0$, and therefore
$$
\Delta = \int^{q_s}_{q_0}\!\! dq\,\, \frac{H_{qq} }{2H_q}=
\frac{1}{2}\ln \left| \frac{V''(q_s) D(q_s)}{ V''(q_0) D(q_0)}
\right| \,,$$ where we have made use of $V'(q_0)=V'(q_s)=0$. Using
this equality again one finds  $S''(q_{0,s})=
V''(q_{0,s})/D(q_{0,s})$, and finally, plugging all together in
Eq.~(\ref{tau1}), one obtains
\begin{equation}\label{taumn}
\tau=\frac{2\pi}{\sqrt{V''(q_0)|V''(q_s)|}}\,\sqrt{\frac{D(q_s)}{D(q_0)}}\,\,\, e^{\int_{q_0}^{q_s}\!dq\, V'(q)/D(q)},
\end{equation}
in agreement with previous calculations~\cite{hanggi}. Assuming a constant diffusion coefficient $D(q)=T$ (i.e. additive noise), one recovers Kramers result \cite{kramers}.
%\begin{equation}\label{Kramers}
%    \tau = \frac{2\pi}{\sqrt{|V''(q_s)|V''(q_0)}}\, \,\, e^{(V(q_s) -V(q_0))/T}\,,
% \end{equation}
Notice that the role of $N$ is played by $1/T$.

{\em Higher moments of noise.} Consider now Kramers problem of a dissipative particle subject to a white, {\em non}-Gaussian noise. The corresponding Hamiltonian reads as
\begin{equation}\label{HON}
H(q,p) = \epsilon_k p^k + T p^2 -p V'(q)\,.
\end{equation}
Here $k=3,4,\cdots$ and $\epsilon_{3,4,\ldots}$ is the third, fourth, {\em etc} (i.e. first non-vanishing beyond the second) irreducible moment of the noise correlation function. This type of noise appears as e.g. higher order corrections in the Kramers-Moyal expansion of the master equation \cite{gardiner}. Assuming that the higher moments are much smaller than the second one \cite{foot2} and proceeding as in the last case we find
\begin{eqnarray}\nonumber
    \tau = \frac{2\pi}{\sqrt{|V''(q_s)|V''(q_0)}}\, \,\, e^{(V(q_s) -V(q_0))/T} \times \\
    \exp \left\{ -\frac{\epsilon_k}{T^k}\int_{q_0}^{q_s} \left[V'(q) \right]^{k-1} dq + O(\epsilon_k^2) \right\}\, .
 \label{KramersHON}
\end{eqnarray}
As can be seen, the prefactor remains unchanged and the whole
contribution coming from the higher order noise concentrates in an
extra ''phase''. Note that $\epsilon_k$ is necessarily positive
for even $k$ (in order to keep the noise real) but it can be
either positive or negative for odd $k$. For the escape processes
under consideration $V(q_s) > V(q_0)$, and so the integral term in
the extra ''phase'' is positive, what implies that even moments of noises
only contribute to reduce the escape time, while the odd ones
can reduce or increase the switching time, depending on
the sign of the corresponding moment.

To conclude we have calculated the escape rate from a metastable
state whose  dynamics is described by a general multi-step master
equation. We found a relatively simple analytical result for
switching rates between metastable states (but not for absorbing
phase transition, as e.g. extinction) of an arbitrary single-species
reaction scheme. We have shown that the general formula found here
reduces to  known results for  single-step reactions and
Fokker-Planck equations, with either additive or multiplicative
noises.

We are indebted  to M.~Dykman, B.~Meerson and P. Sasorov  for numerous useful
discussions. C. E. is grateful to the William I. Fine Theoretical
Physics Institute for its hospitality. This work has been
partially supported by the MEC (Spain) through Project No.
FIS2005-01729; A.K.  was supported by NSF grants DMR-0405212 and
DMR- 0804266.


\begin{thebibliography} {99}
\bibitem{foot3} In so doing we neglect the back-current from the state centered at $q_0'$, see Fig.~\ref{fig1}, towards $q_0$.
This is appropriate at times shorter than (exponentialy long) escape time from $q_0'$.
\bibitem{foot1} Notice that equation of motion along the activation  trajectory $\dot q=H_p(q,p_a(q)) = w_- -w_+= - H_p(q,0)$ is the time-reversal
partner of the relaxation motion.  This is a consequence of the fact that
the single-step reactions satisfy detailed balance condition. We are
indebted to M. Dykman for discussion of this point.
\bibitem{foot2} Here, weakness is to be understood  in the sense that $O(\epsilon_k^2)$ terms in the exponent may be disregarded.
This is the case when $\epsilon_k^2(k-1)\int_{q_0}^{q_s}[V'(q)]^{2k-3}dq \ll T^{2k-1}$.
\end{thebibliography}
\end{document}